\documentclass[lettersize,journal]{IEEEtran}
\usepackage{amsmath,amsfonts}
\usepackage{algorithmic}
\usepackage{algorithm}
\usepackage{array}
\usepackage[caption=false,font=normalsize,labelfont=sf,textfont=sf]{subfig}
\usepackage{textcomp}
\usepackage{stfloats}
\usepackage{url}
\usepackage{verbatim}
\usepackage{graphicx}
\usepackage{bbding}
\usepackage{booktabs}
\usepackage{multirow}
\usepackage{biblatex} 
\usepackage{makecell}
\usepackage{picins}
\usepackage[marginal]{footmisc}

\begin{document}

\title{MSEVA : A System for Multimodal Short Videos Emotion Visual Analysis}

\author{Qinglan Wei,
Yaqi Zhou,
Longhui Xiao,
Yuan Zhang$^{*}$
\thanks{The authors are with the Communication University of China, Beijing 100000, China (e-mail: qlwei@cuc.edu.cn; yqzhou@cuc.edu.cn; longhuix@126.com; yzhang@cuc.edu.cn)}
\thanks{*: Corresponding author} 

}

\markboth{IEEE TRANSACTIONS ON COMPUTATIONAL SOCIAL SYSTEMS,~Vol.~XX, No.~X, December~2023}%
{Shell \MakeLowercase{\textit{et al.}}: A Sample Article Using IEEEtran.cls for IEEE Journals}


\maketitle

\footnote{First Author and Second Author contribute equally to this work.\\}

\begin{abstract}
YouTube Shorts, a new section launched by YouTube in 2021, is a direct competitor to short video platforms like TikTok. It reflects the rising demand for short video content among online users. Social media platforms are often flooded with short videos that capture different perspectives and emotions on hot events. These videos can go viral and have a significant impact on the public’s mood and views. 
However, short videos’ affective computing was a neglected area of research in the past. Monitoring the public’s emotions through these videos requires a lot of time and effort, which may not be enough to prevent undesirable outcomes. 
In this paper, we create the first multimodal dataset of short video news covering hot events. We also propose an automatic technique for audio segmenting and transcribing. In addition, we improve the accuracy of the multimodal affective computing model by about 4.17$\%$ by optimizing it. Moreover, a novel system MSEVA for emotion analysis of short videos is proposed. 
Achieving good results on the bili-news dataset, the MSEVA system applies the multimodal emotion analysis method in the real world. It is helpful to conduct timely public opinion guidance and stop the spread of negative emotions. 
Data and code from our investigations can be accessed at:  \url{http://xxx.github.com}.
\end{abstract}


\section{Introduction}
\IEEEPARstart{S}{hort} video has become a wide range of production and dissemination of a multimodal media format attributed to their convenience and accessibility. With the rise of mobile internet technology, a variety of short video platforms has opened up a short video era for the audience. The length of a short video is usually measured in seconds. It refers to a new short video that is played on the network platform for people to watch, browse, and share at any time. It spreads to the audience through mobile internet technology, with entertainment, fashion, and opinions about current events as the main content, so as to gain the attention of the audience \cite{yang2019analysis}. \par

Nowadays, short video is one of the most important media formats for the dissemination of hot events or topics. 
\textbf{The main characteristics of short videos are as follows: First,} short videos contain rich modal content such as video, audio, and text, with each modality being crucial for emotion analysis. Therefore, this paper employs the method of multimodal emotion analysis to analyze the emotion of short videos by combining the video, audio, and textual data of short videos.\textbf{ Second,} short videos have high transmission speeds. Due to the rich social interaction functions of the short video platform, users can comment on short videos, forward them, and even create short videos inspired by their favorite content. Consequently, analyzing the inherent emotion of popular short videos related to hot events can help us comprehend public attitudes and anticipate the direction of public opinions.\textbf{ Third,} short videos typically have simple but strong emotions. Because short videos have the characteristics of fragmented transmission, they abandon the form and logic of traditional videos in the past. Instead, short videos are created for strong emotional impacts on the audience in a short duration, in order to gain high likes and comments. Thus, compared to traditional videos, conducting multimodal emotion analysis on short videos can usually get more accurate results, which effectively harnesses the potential of short videos as a burgeoning multimedia resource.\par

\begin{figure}[!t]
    \centering
    \includegraphics[width=0.45\textwidth]{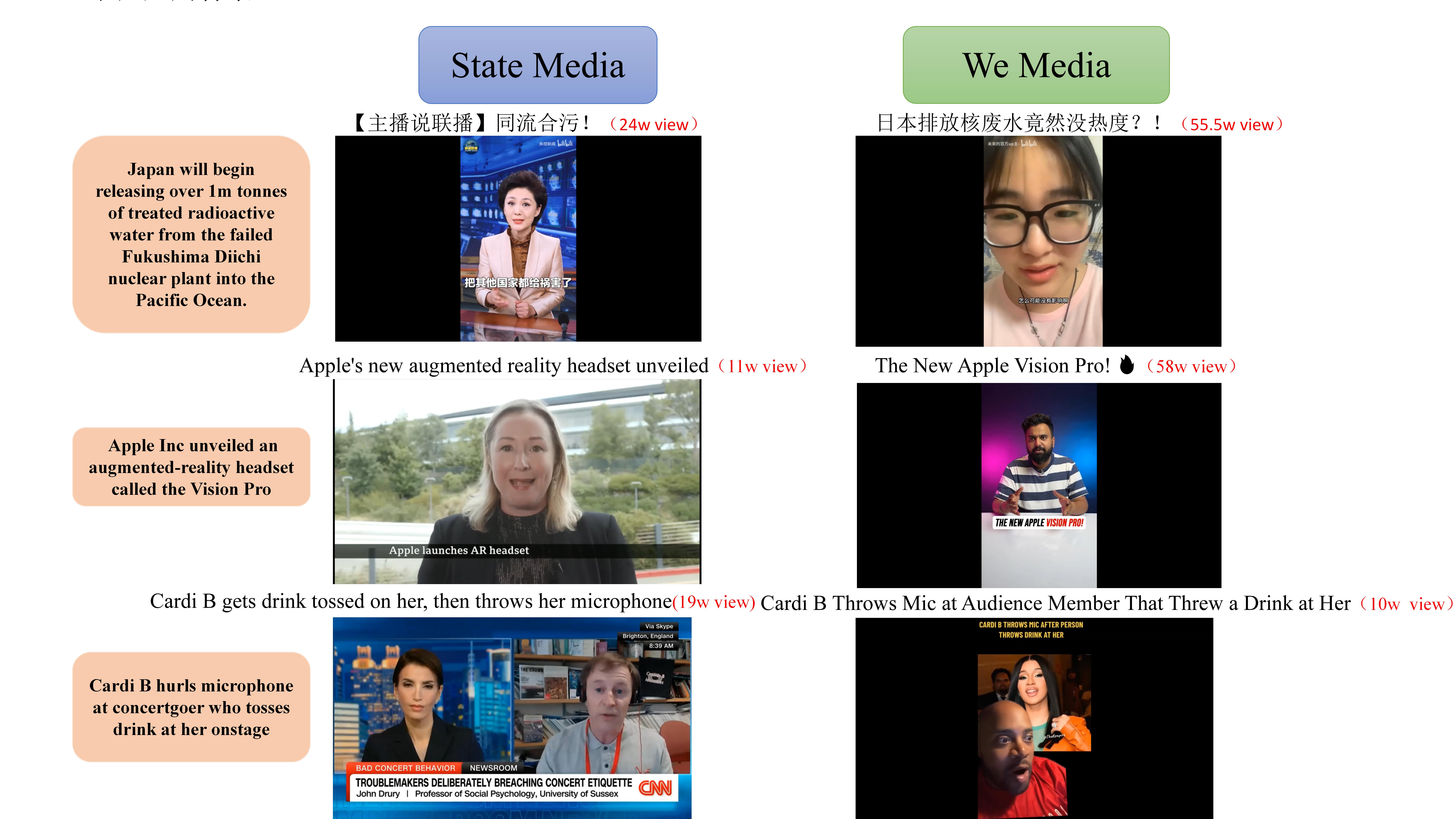}
    \caption{Examples of emotional influence from state media on we media}
    \label{fig1}
\end{figure}

In our analysis of hot events on short video platforms, we observed that the emotions of short videos posted by state media exert significant influence on we media and the public. 
The analysis for different platforms and hot events is shown in Figure \ref{fig1}. Taking the Chinese short video platform Bilibili as an example, in response to Japan's decision to discharge nuclear wastewater from the Fukushima nuclear power plant into the Pacific Ocean, CCTV News posted a short video titled ``Associate with Evil Elements". This short video condemned Japan's action of releasing nuclear wastewater with an angry tone. Subsequently, another short video was posted by Chinese we media with the title ``The Discharge of  Nuclear Wastewater Has Not Been Discussed Yet", urging the Chinese public to recognize the dangers of Japan's nuclear wastewater discharge. These two short videos both have high view counts. 
In short video platforms, the emotional interpretations are popular with the audience, which can be said to keep up with the trend of the times \cite{yang2019analysis}. Such short videos often trigger widespread emotional resonance due to the strong interactivity of internet platforms. Therefore, emotion analysis on short videos is a significant research focus. \textbf{The MSEVA system that we designed can monitor the latent emotions of short videos on short video platforms and regulate public opinion quickly.}\par

\textbf{The key contributions of this work are as follows: }\par
1) A multimodal short video dataset named bili-news is constructed, which has overall annotation of short video emotions. The automatic audio segmentation and transcription method is proposed to improve the efficiency throughout the dataset construction process. Also, the annotation enhances the emotion recognition. The dataset is openly accessible.\par
2) We have improved the accuracy of the result by approximately 4.17$\%$ by optimizing the method of text modality based on the V2EM \cite{wei2022fv2es} multimodal emotion analysis model. Additionally, we conduct experiments considering the comparison of state-of-the-art small and large language models.\par
3) The MSEVA system is a novel emotion analysis approach that we propose in this paper. It is designed for short videos and addresses the research gap discussed in Section \ref{sec:2.2} and \ref{sec:2.3}. The system achieves end-to-end emotion analysis for short videos and provides visualized results, including emotions of comprehensive analysis, emotions of individual modalities, and temporal analysis. The system is open-source.\par

\section{Related Work}
\subsection{Datasets of Multimodal Emotion Analysis}
\label{sec:2.1}
Among current datasets, the dialogs of the IEMOCAP \cite{busso2008iemocap} dataset were manually segmented at the dialog turn level, and the professional transcription was obtained from Ubiqus. All videos in CMU-MOSEAS \cite{zadeh2020cmu} have manual and punctuated transcriptions. Punctuation markers are used to separate sentences, similar to CMU-MOSEI \cite{zadeh2018multimodal}. In the process of dataset construction, we found that current methods mainly rely on manual segmentation and transcription, as shown in Table \ref{table1}. Although this method ensures dataset accuracy and richness, it has some limitations. The manual segmentation and transcription processes need substantial human effort and time, hindering the update of the dataset. Besides, the utterance level segmentation and annotation might overlook the overall emotions of short videos, which is a focal point of our study.\par 
The current datasets for multimodal emotion analysis are composed of segments of long videos. However, the short video is an independent and complete media form that differs from segments of long videos. We need to focus on the short videos on social media platforms.\par
\textbf{The construction of datasets currently depends on the manual segmentation and annotation method, which requires a lot of human effort. In addition, most datasets consist of long videos with emotional annotations for each utterance. Hence, we need to build a dataset for short videos with emotional annotations for the whole video.}\par

\begin{table}
\renewcommand\arraystretch{1.3}
        \centering
	\caption{Segmentation and transcription methods of current multimodal dataset}
	\label{table1}
	\begin{center}
		\begin{tabular}{p{2.3cm}<{\centering} p{1.2cm}<{\centering} p{1.2cm}<{\centering} p{1.2cm}<{\centering} p{1.2cm}<{\centering}}
			\toprule  
 	\textbf{Dataset}&\textbf{\makecell[c]{Manual\\Segment-\\ation}}&\textbf{\makecell[c]{Auto-\\matic\\Segment-\\ation}}&\textbf{\makecell[c]{Manual\\Trans-\\cription}}&\textbf{\makecell[c]{Auto-\\matic\\Trans-\\cription}}  \\
			\midrule[0.5pt]
			IEMOCAP \cite{busso2008iemocap} & \Checkmark & \XSolid & \XSolid &  \Checkmark  \\
			CMU-MOSI \cite{zadeh2015micro} & \Checkmark & \XSolid &\Checkmark & \XSolid  \\ 
            CMU-MOSEI \cite{zadeh2018multimodal} &\Checkmark & \XSolid &\Checkmark & \XSolid  \\ 
            UR-FUNNY \cite{hasan2019ur}& \Checkmark & \XSolid & \Checkmark & \XSolid  \\ 
            CH-SIMS \cite{yu2020ch} & \Checkmark & \XSolid & \Checkmark & \XSolid  \\ 
            CMU-MOSEAS \cite{zadeh2020cmu} &\Checkmark & \XSolid & \Checkmark & \XSolid  \\ 
			\bottomrule
		\end{tabular}
	\end{center}
\end{table}

\subsection{General Multimodal Analysis of Short Videos}
\label{sec:2.2}
Early research mainly focused on short videos on the Vine platform. In 2014, Redi et al. \cite{redi20146} proposed a set of computational features like audio features and visual features that they map to the components of creativity and a supervised approach to automatically detect creative videos. In 2016, Zhang et al. \cite{zhang2016shorter} proposed a tree-guided multi-task multi-modal learning model to estimate the venue category for each unseen microvideo. In the same year, Chen et al. \cite{chen2016micro} proposed a TMALL model for popularity prediction, which was the earliest prediction analysis on the popularity of short videos. In 2020, for Tiktok and MovieLens, two micro-video recommendation datasets, Tao et al. \cite{tao2020mgat} developed a new method MGAT, which incorporates attention mechanism into the graph neural network framework, to disentangle user preferences on different modalities. In 2023, Qi et al. \cite{qi2023fakesv} constructed FakeSV, the largest short video dataset about fake news based on  Douyin and Kuaishou, and provided a new multimodal detection model SV-FEND which exploits the cross-modal correlations to select the most informative features and utilizes the social context information for detection. \par
 \textbf{The analysis of multimodal data of short videos is a hot research topic, involving areas such as popularity prediction, location classification, video recommendations, and fake video detection. However, current studies only focus on objective features and their relation with user behavior, neglecting the intrinsic emotion.}\par

\subsection{General Multimodal Emotion Analysis of Videos}
\label{sec:2.3}
Early research on video emotion analysis was not in the wild, their research primarily focused on movie segments and movie review data. In 2003, Kang et al. \cite{kang2003affective} discussed a new technique for detecting affective events using Hidden Markov Models (HMM), based on low-level features, including color, motion, and shot cut rate. In 2006, Wang et al. \cite{wang2006affective} proposed the combination of visual and audio features with support vector machines and achieved good results. In 2013, with the rapid development of multimedia social platforms, Wollmer et al. \cite{wollmer2013youtube} focused on automatically analyzing a speaker's sentiment in online videos containing movie reviews. In addition to textual information, this approach considers adding audio features as typically used in speech-based emotion recognition as well as video features encoding valuable valence information conveyed by the speaker. In 2018, in order to process a large number of online videos and improve the processing power of real-time emotion analysis, Tran et al. \cite{tran2018ensemble} proposed a real-time multimodal emotion analysis model, which leveraged the processing speed of extreme learning machine and graphics processing unit to overcome the limitations of standard learning algorithms and central processing unit (CPU).\par
There is some research that take GIFs as objects of emotion analysis, which are similar to our study. However, these GIFs consist of only a few frames, which is quite different from short videos. In 2014, Jou et al. \cite{jou2014predicting} proposed the first model to predict the emotions perceived by viewers after they are shown animated GIF images. In 2019, Yang et al. \cite{yang2019human} proposed KAVAN network which consists of a facial attention module and a hierarchical segment temporal module to conduct human-centered GIF emotion recognition.\par
As for the end-to-end video emotion analysis methods, there are still few relevant studies. General existing works on multimodal emotion analysis adopt a two pipeline, first extracting feature representations for each single modality and then performing end-to-end learning with the extracted feature. In 2020, Zhao et al. \cite{zhao2020end} proposed to recognize video emotions in an end-to-end manner based on convolutional neural networks (CNNs), and developed a deep Visual-Audio Attention Network (VAANet). In 2021, Dai et al. \cite{dai2021multimodal} developed a fully end-to-end model FE2E that connects the two phases and optimizes them jointly. In 2022, Wei et al. \cite{wei2022fv2es} designed a fully multimodal video-to-emotion system FV2ES for fast yet effective recognition inference. For visual modality, FV2ES used RepVGG to improve the efficiency of multi-modal emotion analysis. The Hierarchical-Attention Spectrum Computing Module was used to improve the accuracy of the model for audio modality, and the pre-trained Albert model was used for feature extraction and prediction for textual modality.\par
 {\bf{Earlier research on video emotion analysis mostly concentrated on movie segments and review data, and the end-to-end emotion analysis was limited. Even though there were some multimodal affective computing methods, they were not suitable for short videos.}} \par

\section{Our Work}
\subsection{Bili-News Dataset Construction}
\label{sec:1}
As discussed in Section \ref{sec:2.1}, automatic utterance-level segmentation and transcription methods have not been adopted in current multimodal emotion analysis datasets. Most existing datasets focus on emotion for utterance-segmented videos, lacking overall annotations for the emotions of the entire short videos. {\textbf{In this section, we present the bili-news dataset construction, which involves two steps: (a) employing automatic segmentation and transcription methods and (b) selecting and assigning overall emotion annotations for the dataset.}} The following subsections describe the process of constructing this dataset in more detail.\par
\begin{figure}[!t]
    \centering
    \includegraphics[width=0.45\textwidth]{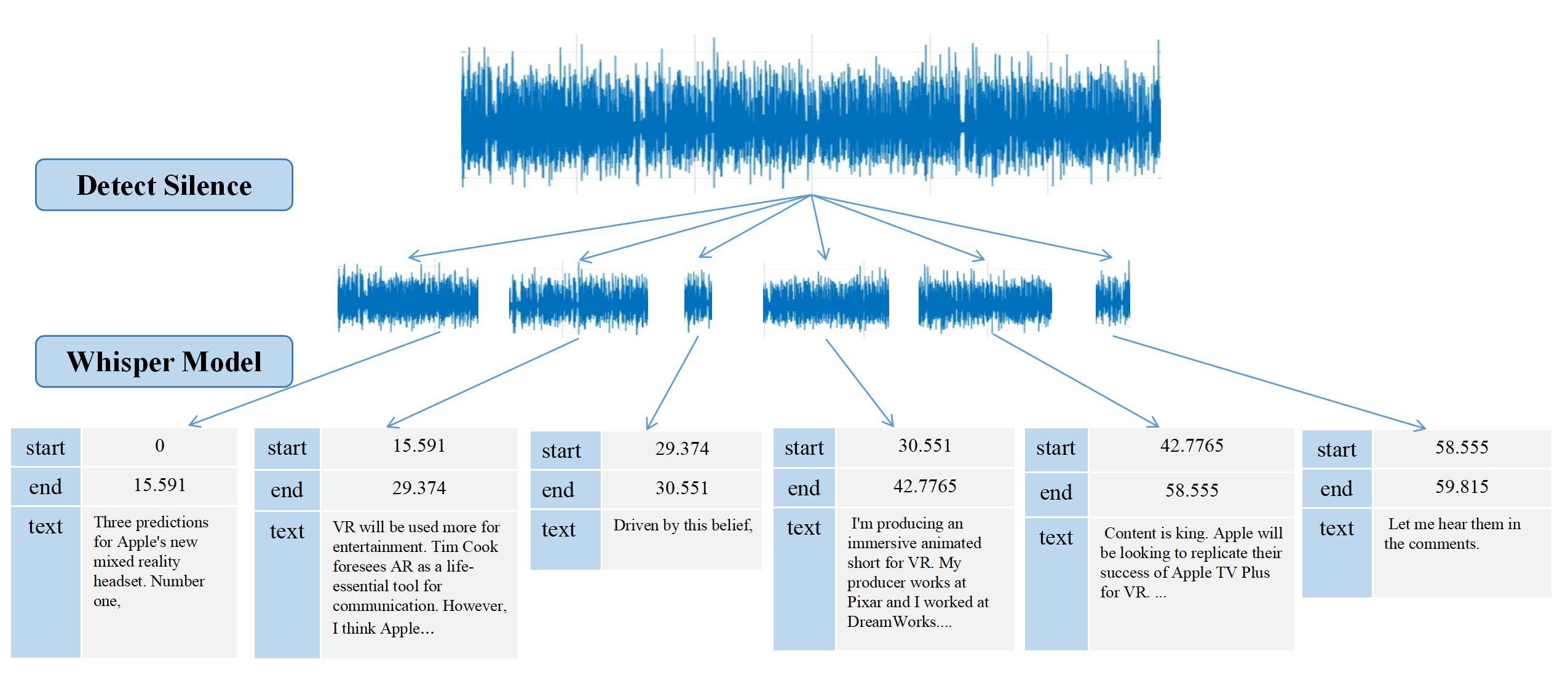}
    \caption{The process of automatic segmentation and transcription method}
    \label{fig2}
\end{figure}

1) \textit{Automatic Segmentation and Transcription Method}
\ 
\newline
\indent
In this section, we propose the first automatic segmentation and transcription method and use it in the process of bili-news construction. \textbf{According to the speaker’s speech rhythm, we segment the audio part of short videos and obtain the start time and end time of each sentence. We then feed the audio segments to the Whisper model \cite{radford2023robust} which transcribes the speech into English text in a consistent way.} The process is shown in Figure \ref{fig2}. This method greatly reduces the cost of manual segmentation and transcription and enhances the efficiency of dataset construction.\par
The detect-silence function in pydub library is used to detect the silence interval in speech. According to our experiments, a threshold of 0.8 seconds was selected as the cutoff for segmenting the original audio into short segments corresponding to each sentence. Subsequently, for each short segment, the Whisper model is utilized for speech recognition and translation, generating the subtitle text of each sentence. The segmentation timestamps and subtitle texts were then outputted into files. Since the Whisper model has the available pre-trained optimal model that can be directly utilized. According to the universality of speech recognition and translation tasks, there is no need to add datasets for fine-tuning in practical applications, so this paper will not train the Whisper model additionally. Moreover, the Whisper model supports multi-language speech recognition and translation tasks, such as Chinese$\rightarrow$Chinese, English$\rightarrow$English, Chinese$\rightarrow$English, Korean$\rightarrow$English, and so on, which enables the automatic utterance-level segmentation and transcription of audio in multiple languages.\par

2) \textit{Selecting and Assigning Emotion Annotations}
\ 
\newline
\indent
\textbf{Firstly,} we crawled 1820 short videos related to recent hot events from the Bilibili platform. \textbf{Secondly,} we designed special criteria for our research and manually selected the short videos, resulting in a final set of 165 videos. \textbf{Thirdly,} we invited 12 crowdsourced judges to annotate the emotion of the entire short video in the bili-news dataset. \textbf{Then,} we dropped short videos with unclear emotion annotations, which may not be significant in our research. We ultimately retained 147 short videos and validated the consistency of the labeling dataset. The details of the selection process and the short video subjective annotation experiment are as follows. \par
To ensure the short videos in the bili-news dataset meet our research, we designed some criteria for selecting the short videos we crawled: (a) featuring one or two main characters; (b) the speech is clear and in the same language; (c) the duration is less than three minutes; (d) having a simple and strong emotion. Additionally, we dropped the policy-related short videos to ensure the objectivity of the dataset.\par
For the short videos we selected, we organized a short video subjective evaluation experiment to label the emotion of short videos. In order to control the quality of annotation, a qualification test is designed for the judges. Through the test, the judges who have the habit of browsing short videos and can clearly judge the emotion of short videos are selected. For some English short videos of the dataset, the judges with good scores in CET-4 and CET-6 are specially selected to annotate these videos.\, 
Our experiment included 12 crowdsourced judges (6 men and 6 women). Each short video was randomly assigned to a group of 3 people to annotate for negative, positive, and uncertain labels. In order to ensure the effectiveness of annotation, this paper provides training before the experiment to help judges better distinguish positive and negative emotions. This training introduces the Positive and Negative Affect Schedule (PANAS) of psychology. After learning 20 different specific descriptions of positive and negative emotions, the judges annotated the positive and negative intensity of emotions in short videos. This paper selects the most one among the three annotations to label short videos. Only when at least two annotators agree with the same exact emotion, the short video annotation is considered valid. Finally, 147 short videos are retained in the dataset.\par
To measure the annotation consistency among different judges, this paper calculates the Fleiss 'kappa of the labels of 3 judges in the constructed bili-news dataset, then obtains K$>$0.65, which has a considerable degree of consistency in the annotations. In addition, in order to verify the quality of annotations, this section selected annotated short videos with different annotations which may be confusing, and invited a new judge to annotate selected short videos, and 96$\%$ of the annotations were the same as the original labels. According to the new annotations, we also calculated Cohen's Kappa to measure the consistency with the original annotations, then the result was K$>$0.85. This shows good consistency which proves that the bili-news dataset is available.\par

\subsection{Optimize Multimodal Emotion Analysis Model}
\label{sec:2}
In this section, we proposed a more effective multimodal emotion analysis model V2EM-RoBERTa based on the V2EM model \cite{wei2022fv2es} by optimizing the method of text modality. \textbf{We investigated recent multimodal affective computing models. Then, we performed some experiments with small language models that are commonly used. Moreover, we employed state-of-the-art large language models for text modality inference. We then contrasted the results of the experiments using small and large language models respectively.} The details of our work are as follows.\par
In this section, we conducted experiments and optimizations on its textual modality approach. The reason for selecting the textual modality is shown in Table \ref{table2}, where we show multimodal emotion analysis developed over the past three years. In the ``Modality" column, it is evident that almost all recent models have integrated visual modality (V), textual modality (T), and acoustic modality (A) for comprehensive analysis. In the ``Effect" column, we can see that almost the textual modality has the greatest impact, so we optimize our approach for the textual modality to maximize the performance of the model.\par
From Table \ref{table2}, we can find that most methods in textual modality use pre-trained models of BERT-based models for textual feature extraction. Therefore, we explored various BERT-based models and some small language models for textual features in our experiments, as detailed in Chapter 4. Among them, the RoBERTa model \cite{liu2019roberta} has a larger number of model parameters, uses a larger batch size during training, and uses more datasets including CC-News for training, so it has superior performance in our experiments. \par

\begin{table}
\renewcommand\arraystretch{1.3}
        \centering
	\caption{Summary of current multimodal emotion analysis model}
	\label{table2}
	\begin{center}
		\begin{tabular}{p{1cm}<{\centering} p{2.3cm}<{\centering} p{1cm}<{\centering} p{1cm}<{\centering} p{1.2cm}<{\centering}}
			\toprule  
			 \textbf{Years} & \textbf{Method} & \textbf{Modality} & \textbf{Effect} &{\textbf{\makecell[c]{Textual\\ Model}}}  \\
			\midrule[0.5pt]
			2021 & CMCN \cite{peng2021cross} & V+T & T$>$V & BERT  \\
			2021 & FE2E \cite{dai2021multimodal} & V+T+A & T$>$V$>$A & Transformer  \\ 
            2021 & HFU-BERT \cite{lee2021multimodal} & V+T+A & T$>$A$>$V & BERT  \\ 
            2022 & AMOA \cite{li2022amoa} & V+A+T & T$>$A$>$V & BERT  \\ 
            2022 & CERS \cite{ray2022multimodal} & V+A+T & T$>$A$>$V& BART \\ 
            2022 & FV2ES \cite{wei2022fv2es} & V+A+T & - & ALBERT  \\ 
            2023 & QAP \cite{li2023qap} & V+A+T & T$>$V$>$A & ALBERT\\
            2023 & TETFN \cite{wang2023tetfn} & V+A+T & T$>$V$>$A & BERT \\
			\bottomrule
		\end{tabular}
	\end{center}
\end{table}
\begin{figure}[!t]
    \centering
    \includegraphics[width=0.45\textwidth]{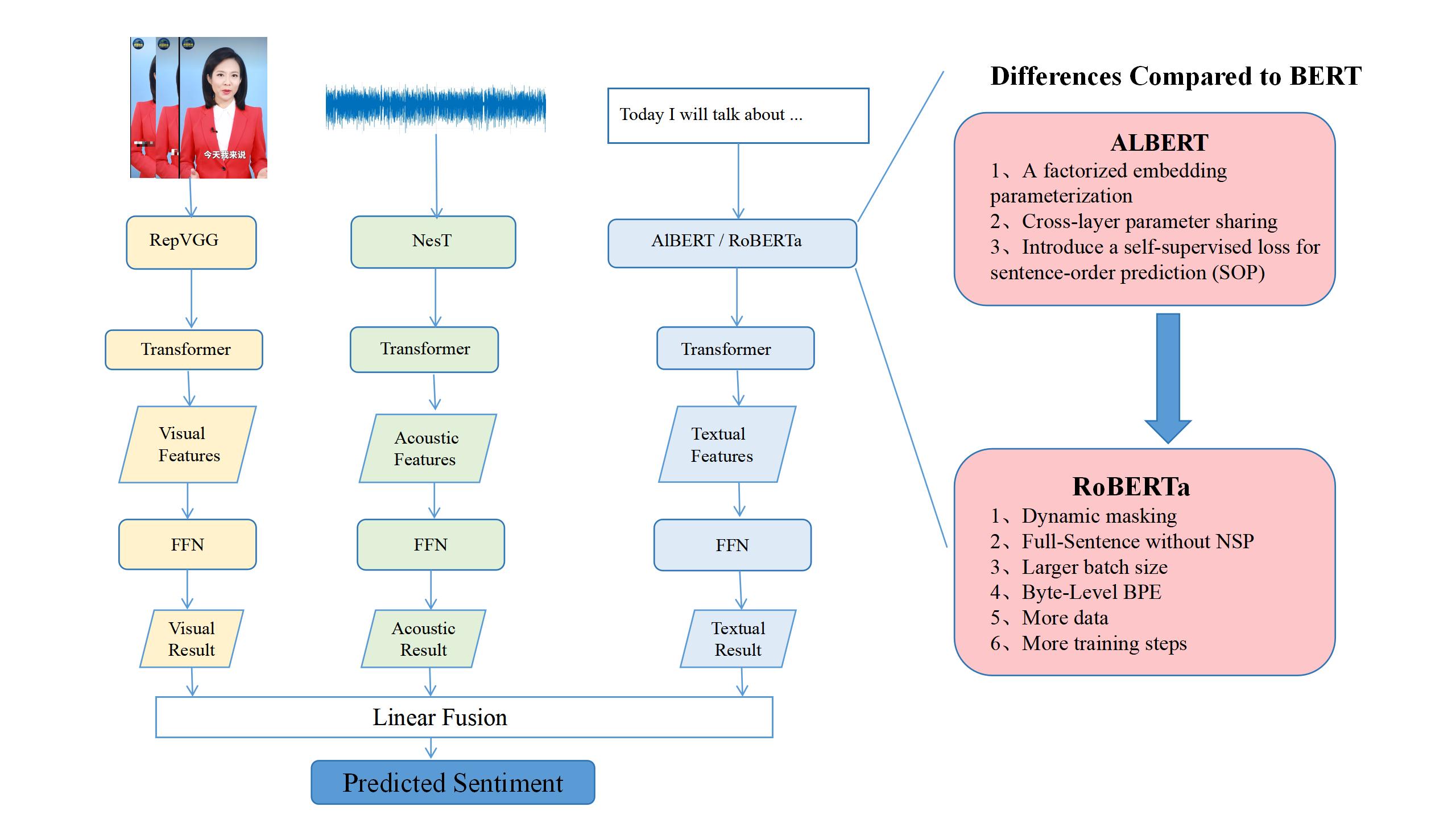}
    \caption{The architecture of V2EM-Roberta multimodal emotion analysis model}
    \label{fig3}
\end{figure}

Recently, large language models have become very popular. Considering the partial similarity between the tasks of text emotion analysis and multimodal emotion analysis, we attempted to employ large language models for textual modality emotion analysis. Subsequently, we combined the results from three modalities through linear fusion to get the final prediction. Considering the size of the number of parameters, we select a large language model with the number of parameters ranging from 200M to 500M for comparative experiments. The experimental results are shown in Chapter 4. The results indicated that the performance of large language models was not as good as small language models trained on the dataset, which verifies the conclusion drawn by Zhang et al. \cite{zhang2023sentiment} that LLMs lag behind in more complex tasks requiring deeper understanding or structured sentiment information.\par
Therefore, based on the open-source end-to-end V2EM model \cite{wei2022fv2es}, we proposed a more effective multi-modal emotion analysis model named V2EM-RoBERTa. For the visual modality, the V2EM-RoBERTa takes the capture of image frames at fixed intervals as input. Due to short videos containing explicit subjects, facial expression is the most important for emotion analysis of a video frame. Indeed, the mtcnn face recognition model is used to intercept the face part of the video frame, and then the RepVGG network is used to extract visual features. The visual features are encoded using a Transformer model with a location-embedded layer containing mode time information. For the acoustic modality, the V2EM-RoBERTa model will extract the log-mel frequency feature of the original audio, expand it into two-dimensional frequency feature graphs,  divide the feature graph into 16 sub-graph sequences, and input them into NesT structure to extract acoustic features. Then the features were inputted into the Transformer model which can model time information to encode the data. For the text modality, we extract the features of the text using the pretrained small language model RoBERTa, then we use the Transformer model to extract the temporal features of the text. Finally, the data of all modalities are input into the Forward Feed Network and get the prediction of all modalities, and we used the linear fusion to get the final prediction. The architecture of the multimodal emotion analysis model in this paper is shown in Figure \ref{fig3}.\par

\subsection{The Construction of the MSEVA System}
\begin{figure*}[!t]
    \centering
    \includegraphics[width=0.9\textwidth]{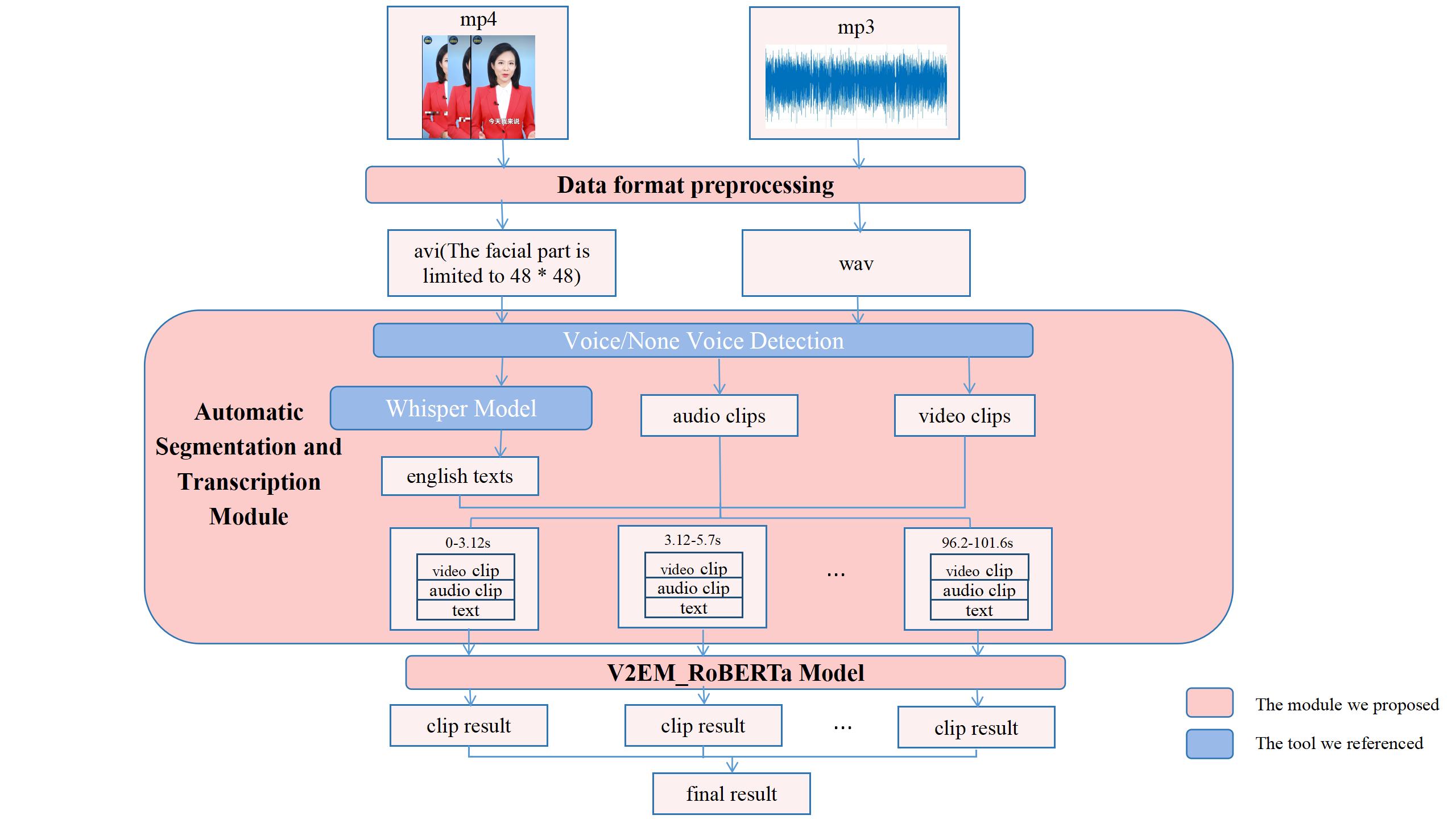}
    \caption{The architecture of multimodal short videos emotion visual analysis (MSEVA) System}
    \label{fig4}
\end{figure*}

The main flow and components of the MSEVA system are shown in Figure \ref{fig4}. \textbf{The system has three main modules}: (a) \textbf{Data Format Preprocessing Module}: this module transforms the short video file that users provide to enable adaptable handling of short videos with various resolutions. (b) \textbf{Automatic Segmentation and Transcription Module}: the module is designed according to the method proposed in Section \ref{sec:1} for utterance-level segmentation and transcription. (c) \textbf{the Pre-trained multimodal emotion analysis Model (V2EM-Roberta)}: the aligned modalities after segmentation are fed into this module to obtain the final result.\par

In the experiment, we found that some short videos with long durations caused substantial memory occupation, but there was no increase in accuracy. To address this problem and finer emotion analysis for short videos, we developed an automatic segmentation and transcription module based on the method in Section \ref{sec:1}. This module generates text files containing the start and end timestamps of each sentence along with the corresponding subtitle text. Then, the segmented audio and video, together with the subtitle text are input into the V2EM-Roberta model for multimodal emotion analysis.\par
\begin{figure}[!t]
    \centering
    \includegraphics[width=0.25\textwidth]{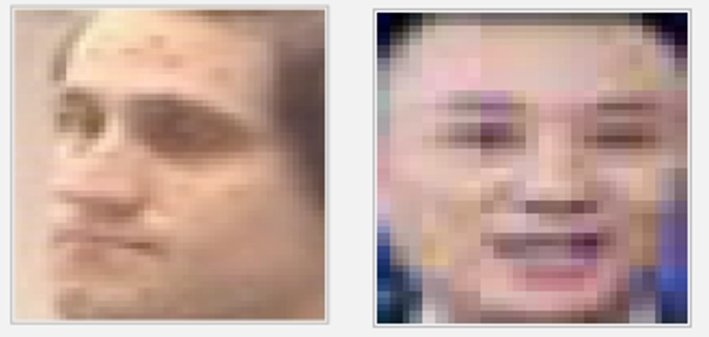}
    \caption{The example for the similar resolution of face area after our data format preprocessing module (the left image is the input during training of IEMOCAP dataset, and the right image is the input from short video during inference of Bili-news dataset)}
    \label{fig5}
\end{figure}

 
The visual modality processing approach of the V2EM-Roberta model using the RepVGG net, where the input face images are of size 48*48. Because of the various resolutions and diverse sizes of face images in the Bilibili platform's short videos, we need to conduct preprocess operations to standardize the data format. The data format preprocessing module is essential to enable the system's adaptability to short videos with different resolutions. \textbf{Our data format preprocessing module workflow is as follows.} We use the FFmpeg tool to convert mp4 and mp3 to avi and wav. \textbf{Then},  According to the statistics in this section, there are four types of short video resolution in the Bili-news dataset, including both landscape and portrait orientations. After several experiments, we devised a compression strategy for different types of videos, as shown in Table \ref{table3}. The facial detection inputted into the V2EM-Roberta model is shown in Figure \ref{fig5}. The left image is the facial detection from the dataset for model training, while the right image is the facial detection from the short video for model inference. It is evident that our compression strategy ensures comparable face image size during both training and inference.\par

\begin{table}
\renewcommand\arraystretch{1.3}
        \centering
	\caption{The compression strategy of short videos with different resolutions}
	\label{table3}
	\begin{center}
		\begin{tabular}{p{3cm}<{\centering} p{3cm}<{\centering}}
			\toprule  
			 \textbf{Original video resolution} & \textbf{Target video resolution}  \\
			\midrule[0.5pt]
			(470$\sim$490)*(550$\sim$570) & 180*224  \\
			(845$\sim$865)*(470$\sim$490)& 214*120  \\ 
                (470$\sim$490)*(840$\sim$860)&	120*214 \\ 
                (1070$\sim$1090)*(1910$\sim$1930)&	144*216\\ 
			\bottomrule
		\end{tabular}
	\end{center}
\end{table}

\section{Experiments}
\subsection{Statistical Analysis of Bili-news Dataset}
\begin{figure}[!t]
    \centering
    \includegraphics[width=0.45\textwidth]{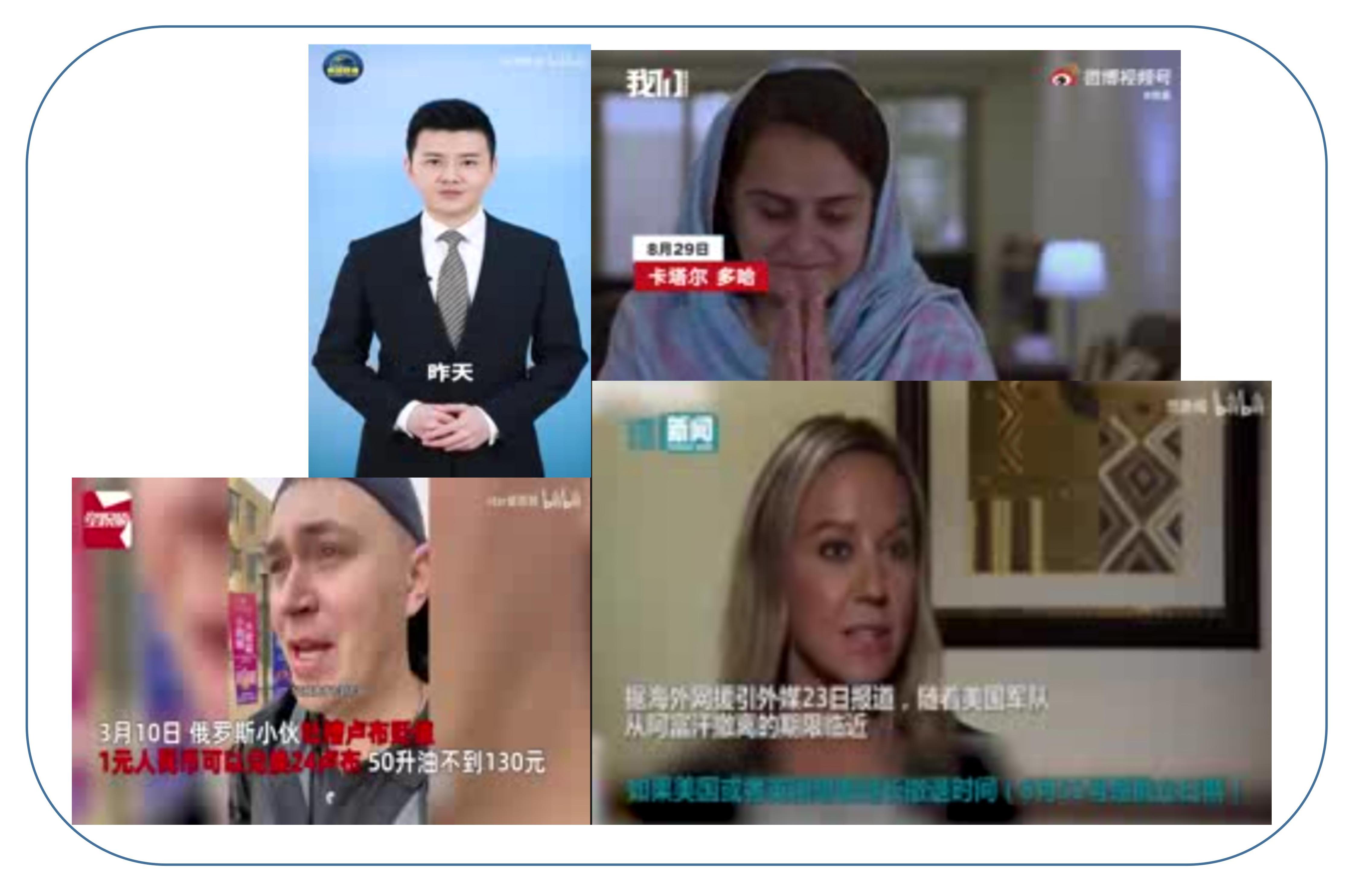}
    \caption{Example snapshots of short videos from our bili-news dataset}
    \label{fig6}
\end{figure}

\textbf{The bili-news dataset has four distinctive characteristics}: (a) \textbf{Definite Emotion}: Short videos have a distinct and strong emotion, with a balanced ratio of positive to negative emotions. (b) \textbf{Diverse Durations}: The dataset has a variety of short video durations. (c) \textbf{Bilingual Content}: Short videos in the bili-news dataset have English and Chinese languages. (d) \textbf{ Various Posters}: Short videos are given by people with different media institutions. The short videos in this dataset show a remarkable diversity, aligning well with the rich variety of short video content typically found on social media platforms, which facilitates objective evaluations. A partial screenshot of the bili-news dataset is shown in Figure \ref{fig6}.\par

In this section, we calculate the emotion annotation of short videos in the bili-news dataset, and the statistical information is shown in column one of Table \ref{table4}. There are 236 positive annotations, 185 negative annotations, and only 20 uncertain annotations, which verify that the short videos in this dataset have clear emotions.\par
Regarding the duration of short videos, if the duration is too short, the video may not effectively convey emotions and might lack the ability to guide or propagate emotions. Conversely, if the duration is excessively long, the video's emotion might change halfway through, potentially conveying positive emotion in the first half and negative emotions in the latter half. Considering the research significance of this paper, we select short videos with durations below three minutes, ensuring that the emotion remains consistent throughout. Among the short videos in the bili-news dataset, there are 40 short videos that lasted less than one minute, 58 short videos that lasted one to one and a half minutes, 37 short videos that lasted one and a half to two minutes, seven short videos that lasted two to two and a half minutes, and five short videos that lasted longer than two and a half minutes. The video duration distribution of the bili-news dataset is shown in column two of Table \ref{table4}. \par
The languages of this dataset are not only Chinese but also English. In the process of dataset construction, we don't restrict the language, except each short video is required to appear in only one language in short videos. Therefore, in the bili-news data set, there are not only 115 short videos in Chinese but also 32 short videos in English. The ratio of Chinese to English is about 4:1, as shown in column three of Table \ref{table4}.\par
The short videos in this dataset are posted by both state media and we media. This dataset encompasses content from 28 prominent Bilibili accounts, including six we media accounts and 22 state media accounts. Specifically, 94 short videos are posted by CCTV News, seven from Phoenix Satellite TV, and six from CGTN. A detailed distribution of the short video poster is shown in column four of Table \ref{table4}.\par

\begin{table*}[pbt]     
\centering                
\caption{Statistical analysis of bili-news dataset}  
\vspace{0.15cm}      
\label{table4}               
\begin{tabular}{|c|c|c|c|}
\hline
\textbf{Label}&\textbf{Duration} & \textbf{Language} & \textbf{Poster} \\ \hline
\begin{minipage}[b]{0.42\columnwidth}
		\raisebox{-.5\height}
        {\includegraphics[width=\linewidth]{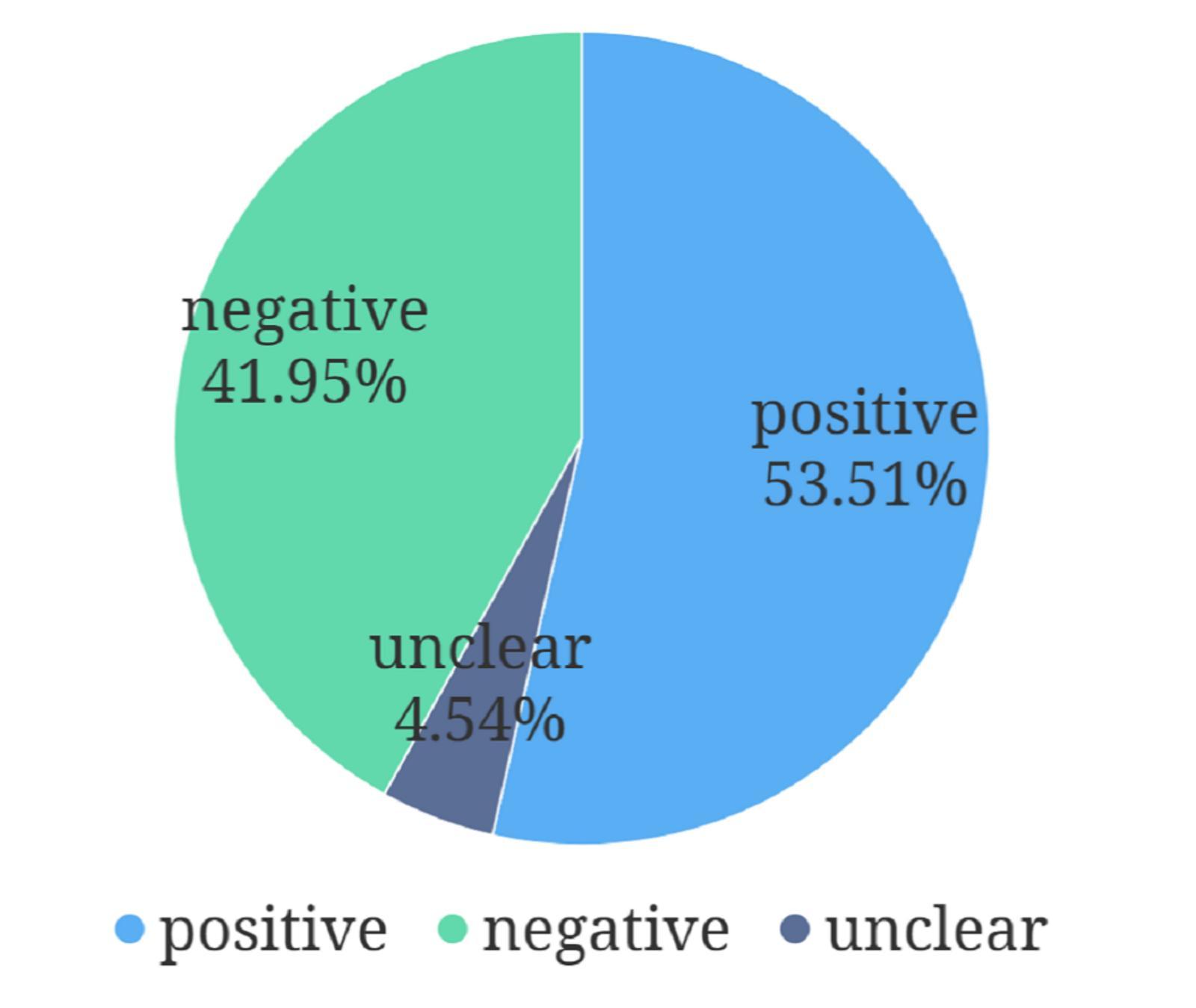}}
	\end{minipage}                    
 &  \begin{minipage}[b]{0.42\columnwidth}
		\raisebox{-.5\height}{\includegraphics[width=\linewidth]{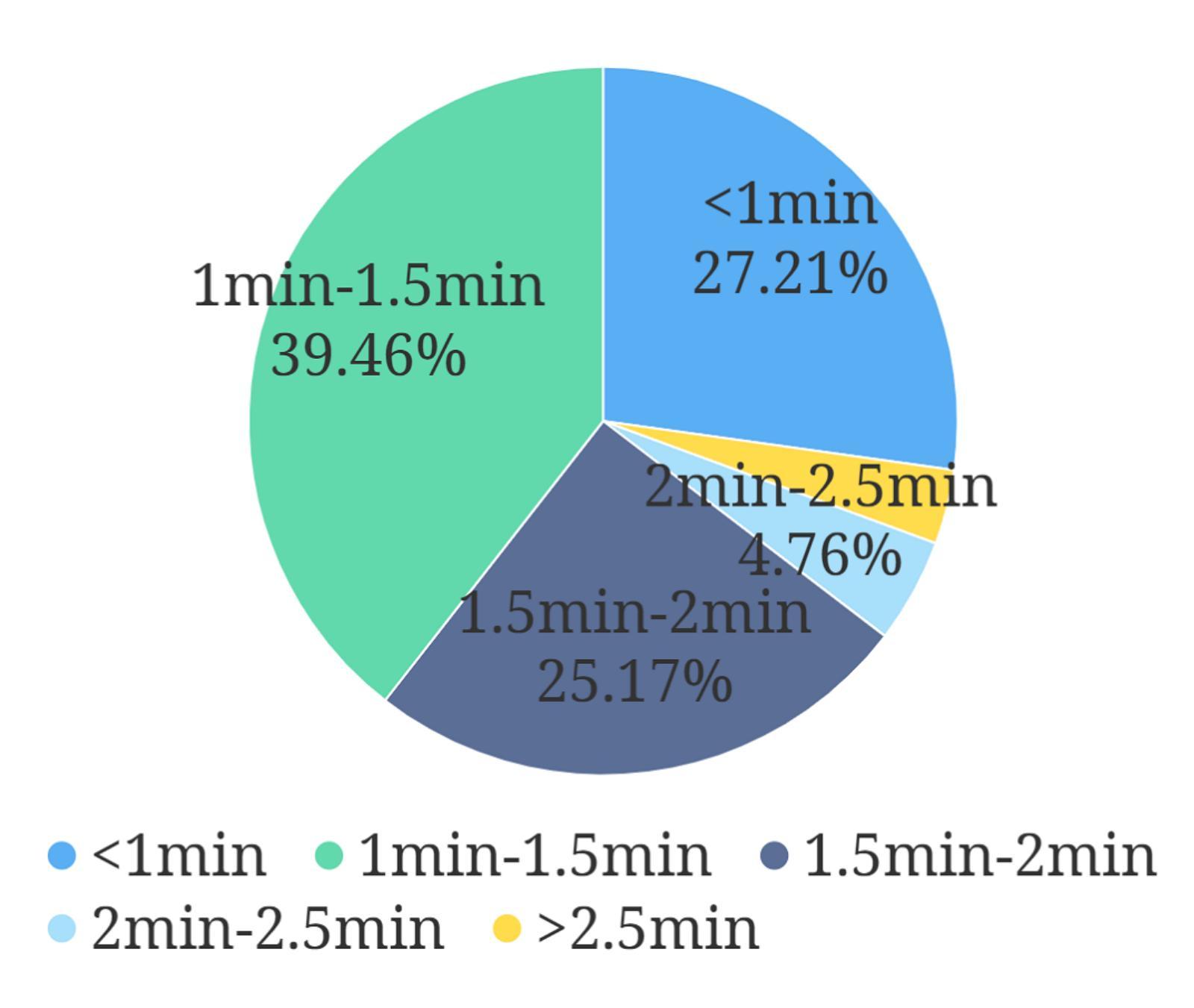}}
	\end{minipage}                       
& \begin{minipage}[b]{0.42\columnwidth}
		\raisebox{-.5\height}{\includegraphics[width=\linewidth]{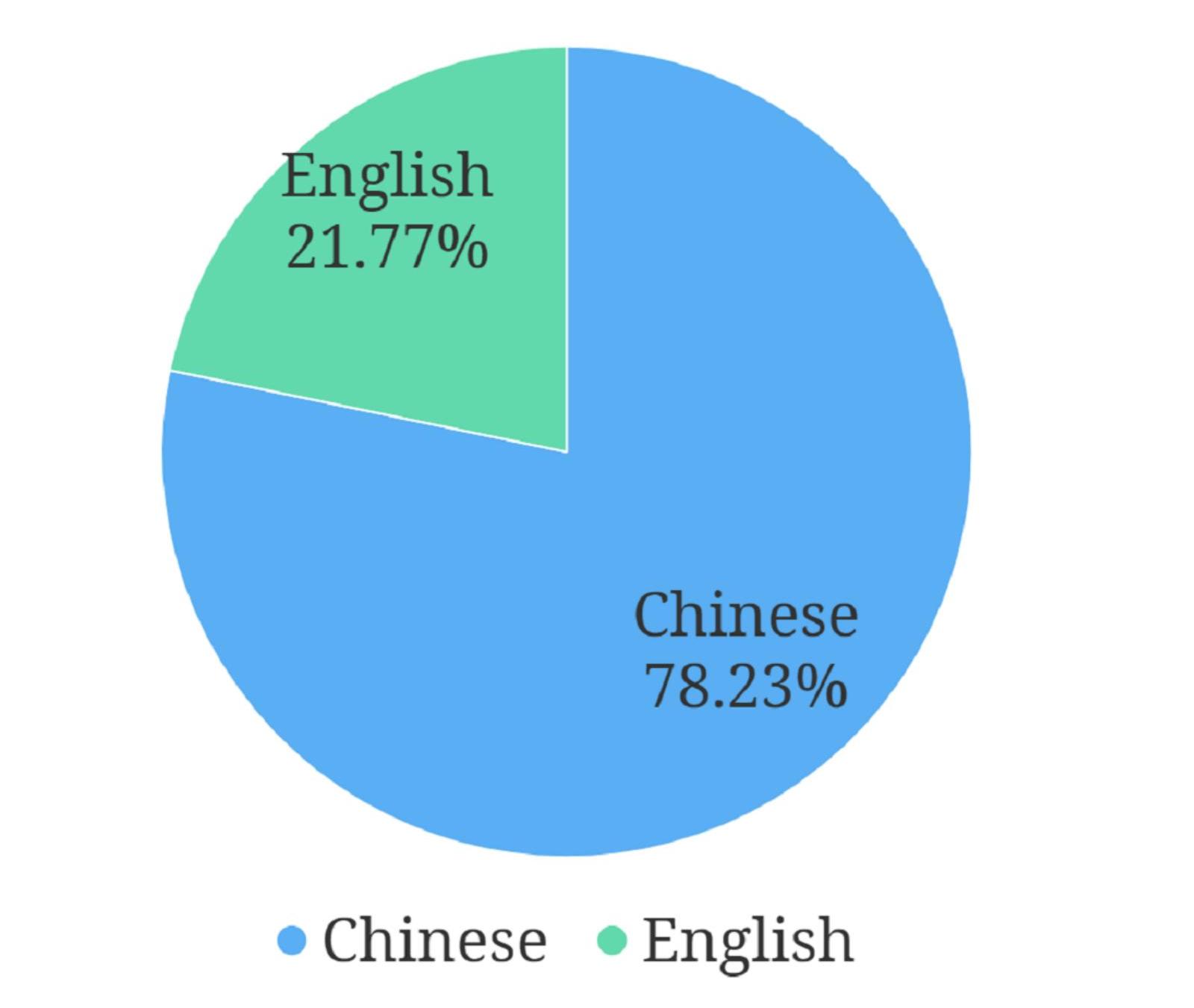}} 
	\end{minipage}                    
 & \begin{minipage}[b]{0.42\columnwidth}
		\raisebox{-.5\height}{\includegraphics[width=\linewidth]{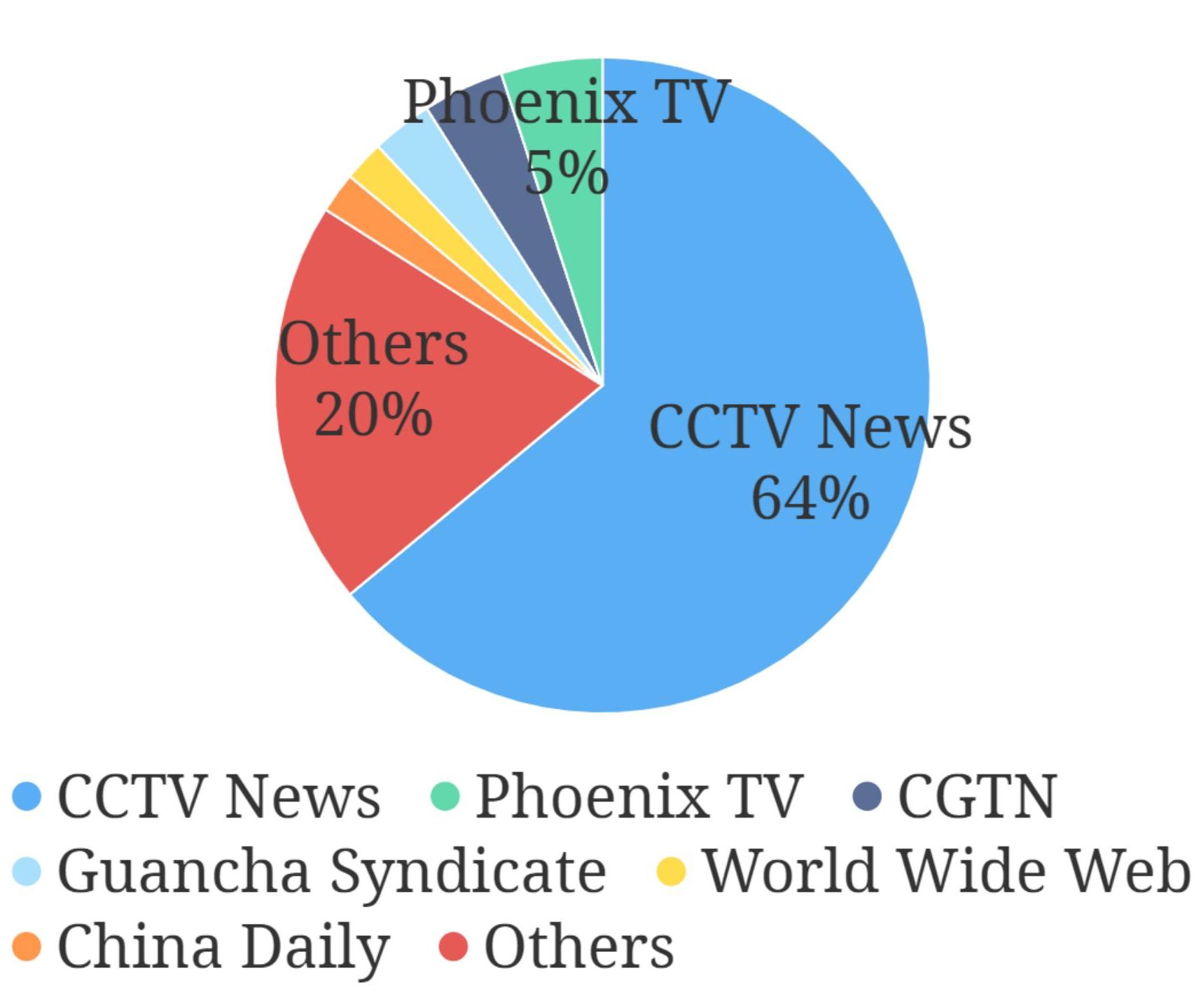}}  
	\end{minipage}                     
 \\ \hline
\end{tabular}
\end{table*}

\subsection{Ablation Study of Automatic Segmentation and Transcription Module}
To validate the necessity of automatic segmentation and transcription modules we introduced in Section \ref{sec:1}, we compared the effects of non-transcribed methods and transcribed methods respectively on the trained multimodal emotion analysis model. In the case of the non-transcribed method, we used the title of the short video crawled as the input of textual modality. We transferred the multimodal emotion analysis model trained on the CMU-MOSEI dataset (label marked -3~-1 as negative and label marked 1-3 as positive) to the bili-news dataset for experiments. The textual inputs for the experiments are the titles we crawled and the subtitles transcribed by our method respectively. \textbf{Using the text generated by this module, instead of using the title, can improve the accuracy of the multimodal emotion analysis model by up to about 10.01$\%$, as evidenced by the experimental results in Table \ref{table5}.}\par

\begin{table}
\renewcommand\arraystretch{1.3}
\setlength\tabcolsep{3pt}
        \centering
	\caption{The result of different textual input (title and transcription)}
	\label{table5}
	\begin{center}
		\begin{tabular}{c|cccc}
			\toprule  
			 \,&\textbf{ACC-2 ($\%$)} & \textbf{Precision ($\%$)}&\textbf{Recall ($\%$)}&\textbf{F1 ($\%$)}  \\
			\midrule[0.5pt]
			\makecell[c]{Non-transcribed\\(Title)} & 74.82 &51.82 &86.36&64.77 \\
			 \textbf{\makecell[c]{Transcribed\\(Transcription)}} & \textbf{82.31} &\textbf{55.37}&\textbf{89.33}&\textbf{68.37}\\ 
			\bottomrule
		\end{tabular}
	\end{center}
\end{table}

In order to validate the effectiveness of the automatic segmentation and transcription method on end-to-end short video multimodal emotion analysis, we compared the performance of subtitles by manually transcribed and subtitles generated by this method on the optimal multimodal emotion analysis model trained on the CMU-MOSEI dataset.These were used as textual modal inputs and tested on the CMU-MOSEI dataset's test set. The experimental results are shown in Table \ref{table6}. In terms of accuracy, there is little difference between manual transcription and automatic transcription. In terms of precision, automatic transcription outperforms manual transcription by 6.63$\%$. However, automatic transcription exhibits lower performance than manual transcription in terms of recall and F1 score. The quality of manual transcription should be higher than that of automatic transcription, so the result of manual transcription is better than automatic transcription. \textbf{Our experiments confirmed that the automatic transcription method has a similar effect to the artificial one in terms of accuracy and precision, and it also saves much labor costs.} \par

\begin{table}
\renewcommand\arraystretch{1.3}
\setlength\tabcolsep{3pt}
        \centering
	\caption{The result of different textual input (manual transcription and automatic transcription)}
	\label{table6}
	\begin{center}
		\begin{tabular}{c|cccc}
			\toprule  
			 \textbf{Transcription}&\textbf{ACC-2 ($\%$)} & \textbf{Precision ($\%$)}&\textbf{Recall ($\%$)}&\textbf{F1 ($\%$)}  \\
			\midrule[0.5pt]
			Manual & 82.08 &78.17 &\textbf{88.81}&\textbf{83.15} \\
			 \textbf{Automatic} & \textbf{82.91} &\textbf{83.35}&82.06&82.70\\ 
			\bottomrule
		\end{tabular}
	\end{center}
\end{table}

\subsection{Performance and Computational Efficiency Analysis of V2EM-RoBERTa Model}
Our experiment relies on CPU Intel (R) Xeon (R) Gold 6326 CPU @ 2.90GHz and GPU Nvidia RTX3090, only one graphics card and CPU are used for training the model, and only the CPU is used for inference.\par
1) \textit{Multimodal Emotion Analysis Experiment}
\ 
\newline
\indent
We conducted optimization experiments for the text modality of the V2EM model on the IEMOCAP dataset \cite{busso2008iemocap} and the CMU-MOSEI dataset \cite{zadeh2018multimodal}. On the IEMOCAP dataset, we extract video frames at a rate of 800 frames per second. The epoch is set to 30, the batch size is set to 1, and the accumulation gradient is set to 4. We combine some small language models for feature extraction of text modality into V2EM models for optimization experiments. As the Section \ref{sec:2} mentioned, we try the pretrained small language models like Albert \cite{lan2019albert}, GPT2 \cite{radford2019language}, BART \cite{lewis2019bart}, Distilbert \cite{sanh2019distilbert}, RoBERTa \cite{liu2019roberta}. The results are shown in Table \ref{table7}. On the MOSEI dataset, due to the long duration of some videos in the dataset and the limitation of the graphics card, we extract a fixed set of 10 video frames per video for the visual modality input. Other experimental parameters remained consistent with those of the IEMOCAP dataset, and the results are shown in Table \ref{table8}. \textbf{We found that using a pre-trained RoBERTa text model \cite{liu2019roberta} improved accuracy by approximately 4.17$\%$ compared to the base model. However, the training time is longer because the number of parameters of the RoBERTa model is larger than that of Albert.}\par

\begin{table*}
\renewcommand\arraystretch{1.3}
        \centering
	\caption{The result of multimodal emotion analysis with different small language models for textual feature extraction based on the V2EM model on the IEMOCAP dataset}
	\label{table7}
	\begin{center}
        \resizebox{\linewidth}{0.1\linewidth}{
		\begin{tabular}{c|ccccccc}
			\toprule  
			 \textbf{Text Model}&\textbf{ACC-2 ($\%$)} & \textbf{Recall ($\%$)}&\textbf{Precision ($\%$)}&\textbf{F1 ($\%$)} &\textbf{AUC ($\%$)}&\textbf{Parameters}&\textbf{Training Time} \\
			\midrule[0.5pt]
			\textbf{Albert-base-v2} & 0.8023 &\textbf{0.6696}&0.4515&0.5335&0.8412&11M&8.89h\\
               \textbf{GPT2} & 0.7508 &0.5527&0.3447&0.4234&0.7416&137M&8.56h\\
			 \textbf{BART} & 0.7833 &0.5958&0.4076&0.4767&0.7951&139M&9.2h\\
            \textbf{distilbert-base-uncased} & 0.8023 &0.5953&0.4504&0.5043&0.8070&67M&8.30h\\
            \textbf{RoBERTa-base} & \textbf{0.8372} &0.6585&\textbf{0.5208}&\textbf{0.5755}&\textbf{0.8587}&125M&9.08h\\
			\bottomrule
		\end{tabular}
        }
	\end{center}
\end{table*}

\begin{table*}
\renewcommand\arraystretch{1.3}
        \centering
	\caption{The result of multimodal emotion analysis experiments with different small language models for textual feature extraction based on the V2EM model on the CMU-MOSEI dataset}
	\label{table8}
	\begin{center}
        \resizebox{\linewidth}{0.1\linewidth}{
		\begin{tabular}{c|ccccccc}
			\toprule  
			 \textbf{Text Model}&\textbf{ACC-2 ($\%$)} & \textbf{Recall ($\%$)}&\textbf{Precision ($\%$)}&\textbf{F1 ($\%$)} &\textbf{AUC ($\%$)}&\textbf{Parameters}&\textbf{Training Time} \\
			\midrule[0.5pt]
			\textbf{Albert-base-v2} &0.7141&0.6137&0.3651&0.4553&0.7254&11M&37.28h\\
               \textbf{GPT2} & 0.6659&0.5617&0.3046&0.3935&0.6538&137M&36.89h\\
			 \textbf{BART} & 0.6995 &0.6088&0.3596&0.4417&\textbf{0.7951}&139M&38.45h\\
            \textbf{distilbert-base-uncased} & 0.7270&0.5686&0.3716&0.4431&0.7187&67M&38.30h\\
            \textbf{RoBERTa-base} & \textbf{0.7328} &\textbf{0.6142}&\textbf{0.3933}&\textbf{0.4722}&0.7437&125M&38.40h\\
			\bottomrule
		\end{tabular}
        }
	\end{center}
\end{table*}

2) \textit{Textual Modal Emotion Analysis Experiment}
\ 
\newline
\indent
In order to validate that RoBERTa model has a better effect than Albert model and other small language models, we conducted a textual modal experiment on the IEMOCAP dataset. The experimental parameters are set the same as the previous experiments, and the experimental results are shown in Table \ref{table9}. \textbf{The results show that the RoBERTa model has the best performance on text modality.}

\begin{table*}
\renewcommand\arraystretch{1.3}
        \centering
	\caption{The result of textual model emotion analysis experiments with different small language models for the textual feature extraction on the IEMOCAP dataset)}
	\label{table9}
	\begin{center}
        \resizebox{\linewidth}{0.1\linewidth}{
		\begin{tabular}{c|ccccccc}
			\toprule  
			 \textbf{Text Model}&\textbf{ACC-2 ($\%$)} & \textbf{Recall ($\%$)}&\textbf{Precision ($\%$)}&\textbf{F1 ($\%$)} &\textbf{AUC ($\%$)}&\textbf{Parameters}&\textbf{Training Time} \\
			\midrule[0.5pt]
			\textbf{Albert-base-v2} &0.8087&0.5681&0.4483&0.4906&0.8027&11M&3.0h\\
               \textbf{GPT2} & 0.6707&0.5906&0.2961&0.3793&0.6908&137M&2.6h\\
			 \textbf{BART} & 0.8083 &0.5758&0.4472&0.4945&0.8168&139M&3.76h\\
            \textbf{distilbert-base-uncased} & 0.8148&\textbf{0.6129}&0.4877&0.5300&0.8382&67M&1.71h\\
            \textbf{RoBERTa-base} & 0.8462 &0.5903&\textbf{0.5354}&\textbf{0.5595}&\textbf{0.8442}&125M&2.78h\\
			\bottomrule
		\end{tabular}
        }
	\end{center}
\end{table*}

3) \textit{Multimodal Experiments Integrated with Large Language Models}
\ 
\newline
\indent
As mentioned in Section \ref{sec:2}, we also conducted experiments using some state-of-the-art large language models like bloomz \cite{muennighoff2022crosslingual}, mt0 \cite{muennighoff2022crosslingual}, flan-t5 \cite{chung2022scaling} on the IEMOCAP dataset. In these experiments, we used a unified prompt command for the pre-trained large language models to inference in the text modality, while other modalities were trained and tested in the same way as the V2EM model. The results are shown in Table \ref{table10}. We can see that using large language models not only takes longer time for training but also the results are not as good as small language models. \textbf{Therefore, for current multimodal emotion analysis tasks, employing small language models proves to be better.}

\begin{table*}
\renewcommand\arraystretch{1.3}
        \centering
	\caption{The result of multimodal emotion analysis experiments integrated with large language models and small language models on the IEMOCAP dataset)}
	\label{table10}
	\begin{center}
        \resizebox{\linewidth}{0.1\linewidth}{
		\begin{tabular}{c|ccccccc}
			\toprule  
			 \textbf{Text Model}&\textbf{ACC-2 ($\%$)} & \textbf{Recall ($\%$)}&\textbf{Precision ($\%$)}&\textbf{F1 ($\%$)} &\textbf{AUC ($\%$)}&\textbf{Parameters}&\textbf{Training Time} \\
			\midrule[0.5pt]
			\textbf{Albert-base-v2 (SLM)} &0.8023&0.6696&0.4515&0.5335&0.8412&11M&8.89h\\
            \textbf{RoBERTa-base (SLM)} & \textbf{0.8372}&\textbf{0.6585}&\textbf{0.5208}&\textbf{0.5755}&\textbf{0.8587}&125M&2.78h\\
            \textbf{bloomz-560m (LLM)} & 0.7530&0.5850&0.3766&0.4497&0.7601&560M&9.53h\\
            \textbf{mt0-base (LLM)} & 0.7411&0.6487&0.3668&0.4612&0.7690&580M&13.07h\\
            \textbf{flan-t5-base (LLM)} & 0.7483&0.5848&0.3531&0.4349&0.7492&248M&11.10h\\
			\bottomrule
		\end{tabular}
        }
	\end{center}
\end{table*}

\subsection{The Test of the MSEVA System Analysis}
\begin{figure}[!t]
    \centering
    \includegraphics[width=0.45\textwidth]{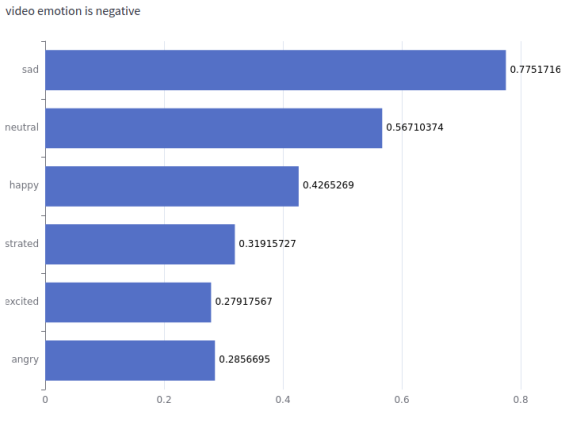}
    \caption{The interface of the multimodal emotion analysis for short videos}
    \label{fig7}
\end{figure}

1) \textit{Comprehensive Emotion Analysis Interface for Short Videos}
\ 
\newline
\indent
\textbf{Clicking the emotion analysis button allows us to do real-time emotion analysis for short videos and make a complete appraisal of emotions.} In this test, we inputted the short video titled ``The female anchor tearfully recalls the fear of the interview" in the bili-news dataset. As shown in Figure \ref{fig7}, the interface first shows the emotion of the video, indicating whether it is positive or negative. Subsequently, it shows specific scores for different emotions, with the highest score being the final result. For this example, the result is sad, which aligns with our subjective judgment and the label.\par

\begin{figure}[!t]
    \centering
    \includegraphics[width=0.45\textwidth]{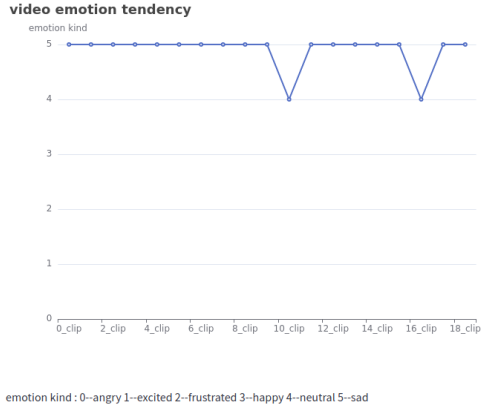}
    \caption{The interface of the temporal emotion analysis for short videos}
    \label{fig8}
\end{figure}

\textbf{To provide finer analysis, our proposed system performs temporal analysis, as emotions in a short video may vary partially throughout its duration.} Leveraging the end-to-end multimodal emotion analysis module based on the pretrianed V2EM-RoBERTa model, we automatically segment the short video into sentences, feeding them into the module to inference for each sentence. We utilize an emotion fluctuation graph for visual representation, enhancing the comprehension of the short video's emotional trajectory. The interface is depicted in Figure \ref{fig8}.  \par

\begin{figure}[!t]
    \centering
    \includegraphics[width=0.45\textwidth]{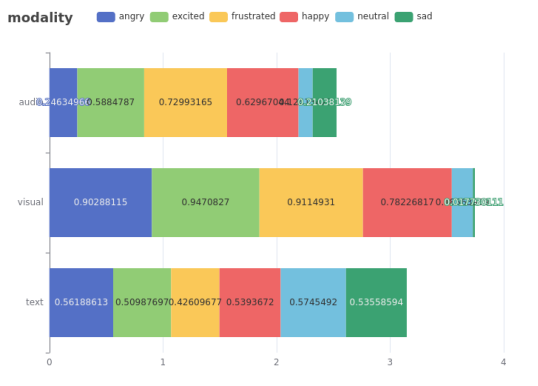}
    \caption{The interface of each modality emotion analysis for short videos}
    \label{fig9}
\end{figure}

\textbf{In addition, the system offers emotion analysis results for each modality, allowing the analysis of short video emotions from various perspectives.} We utilize a decision-level fusion approach in the final result, which means we linearly combine these results of individual modality to get the final result. We show the emotional analysis result for each modality in Figure \ref{fig9}.\par

2) \textit{The Performance of the MSEVA System}
\ 
\newline
\indent
We tested the emotion analysis of the MSEVA system on the  bili-news dataset. The dataset consisted of 147 short videos, including 62 negative videos and 85 positive videos. The result is shown as Table \ref{table11}. \textbf{By comparing the system’s emotion analysis results with the label in the bili-news dataset, we computed the accuracy and F1 score. They were 76.2$\%$ and 81.5$\%$, respectively.}\par

\begin{table}
\renewcommand\arraystretch{1.3}
        \centering
	\caption{The performance of the emotion analysis of the MSEVA system}
	\label{table11}
	\begin{center}
		\begin{tabular}{|c|c|c|}
                \hline
			 \multirow{2}*{Ground Truth} & \multicolumn{2}{|c|}{The result of system}\\
            \cline{2-3}
            ~& Positive & Negative\\
            \hline
            Positive& 77 & 8\\
            \hline
            Negative&27& 35\\
            \hline
		\end{tabular}
	\end{center}
\end{table}
The analysis error of the MSEVA system appears in the case that the subject criticizes and warns against negative behaviors using a humorous way or a lighthearted broadcasting method. In this case, even though the subject's speech contains negative vocabulary such as harm, prohibition, or punishment, the video is still recognized as a positive video by the MSEVA system.\textbf{ When the subject’s speaking style, tone, and content are consistent, the model attains more accurate recognition.}\par 

The wrong case: There is no star when it comes to legal issues, ``Deng Lun" needs to speak and act with caution in life.\par
Case study: The main content of this short video is about the news that the star Deng Lun was fined for tax evasion. Although the anchor's broadcast style is very serious, and the speaking content is also a relatively heavy topic, the audience will have the thought that he deserves his punishment, which is positive, so the analysis result of the model is wrong.\par
The correct case: Ouyang Xiadan: The suspect who beat a 9-year-old boy to death was detained, and the mental problem is not ``immunity". \par
Case study: The short video is mainly about the suspect of the violent incident suffering from mental illness. In view of this social problem, the anchor calls for strengthening the treatment and control of patients with mental illness to prevent them from causing serious social problems. The emotion of the short video is negative, and the result of the MSEVA model is correct.\par
Table \ref{table12} shows the video screenshot of these cases.\par

\begin{table}[pbt]     
\centering                
\caption{Video screenshots of these cases (wrong case on the left, correct case on the right)}  
\vspace{0.15cm}      
\label{table12}               
\begin{tabular}{|c|c|}
\hline
\textbf{Wrong Case}&\textbf{Correct Case} \\ \hline
\begin{minipage}[b]{0.4\columnwidth}
		\raisebox{-.6\height}{\includegraphics[width=\linewidth]{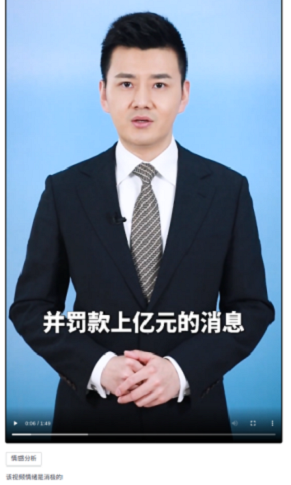}}
	\end{minipage}                    
 &  \begin{minipage}[b]{0.4\columnwidth}
		\raisebox{-.6\height}{\includegraphics[width=\linewidth]{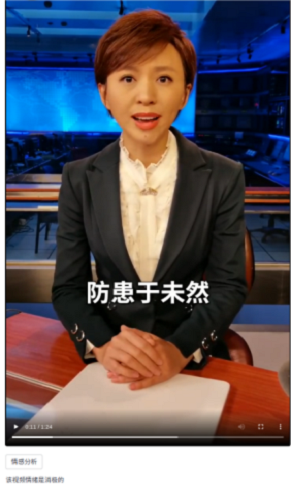}}
	\end{minipage}                                  
 \\ \hline
\end{tabular}
\end{table}

\section{Conclusion and Future Work}
In this paper, we propose the methods of automatic segmentation and transcription method which supports multilingual videos and improve the efficiency of the multimodal dataset construction. This method improves the usability of short video emotion analysis in our life. We firstly construct the multimodal emotion analysis dataset bili-news based on short videos, which includes the annotation for the overall emotion of short videos. This dataset is openly accessible. In addition, we achieved approximately 4.17$\%$ improvement in the accuracy  of the multimodal emotion analysis model based on V2EM \cite{wei2022fv2es}. We conducted several relevant experiments on pre-trained small language models and the current large language models, validating the importance of small language models for multimodal emotion analysis and that large language models cannot completely replace small language models now. Finally, we propose the MSEVA system, designed for end-to-end visual analysis of short videos. This system utilized a multimodel emotion analysis model trained on the CMU-MOSEI dataset and was tested on the bili-news dataset. The results of the experiments show that the system is effective and significant in real-life applications.\par
There are some limitations in this work. The relatively limited number of short videos in the bili-news dataset could be expanded while ensuring data standardization. In the future, we can fine-tune the multimodal emotion analysis model on the expanded bili-news dataset so that we can enhance the model's performance. The performance of the MSVEA system needs to be further improved and to be more available for the real short videos on the platforms. Moreover, the computational time of the current system is high and requires further optimization.\par

\section*{Acknowledgments}
This study was supported by the National Social Science Foundation of China (No. 62301510), the Fundamental Research Funds for the Central Universities (No. CUC23GZ005), the Fundamental Research Funds for the Central Universities (No. CUC23ZDTJ004).

\begin{refcontext}[sorting = none]
\printbibliography
\end{refcontext}

\par\noindent
\parbox[t]{\linewidth}{
\noindent\parpic{\includegraphics[height=1.2in,width=1in,clip,keepaspectratio]{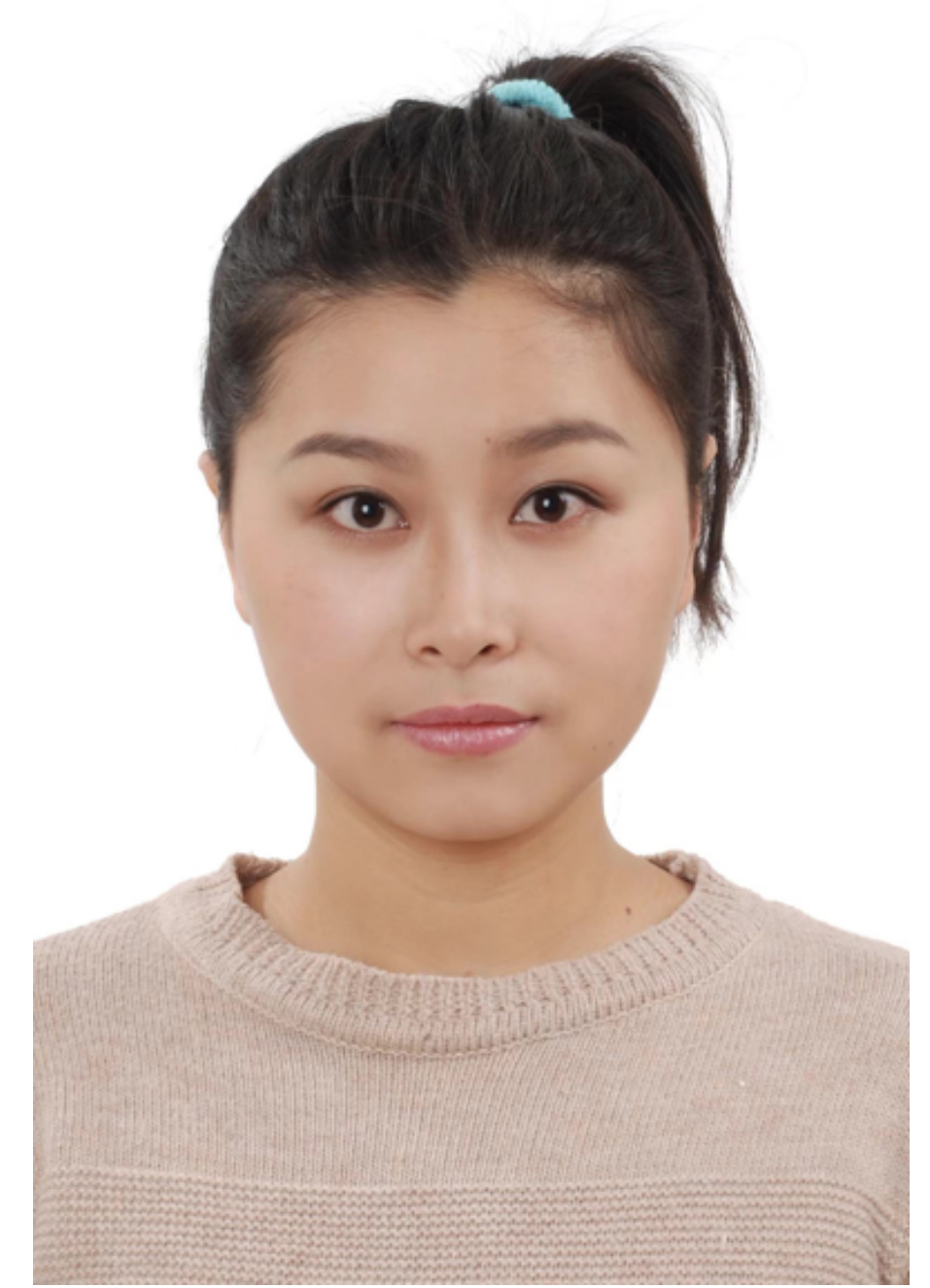}}
\noindent {\bf Qinglan Wei}\ (Member, IEEE) is currently an associate professor with the 
School of Data Science and Intelligent Media, Communication University of China. She was a visiting scholar at the Language Technologies Institute, Carnegie Mellon University, USA. She received the Ph.D. degree in computer science from the College of Artificial Intelligence, Beijing Normal University, China. She has 
published a series of articles in academic journals and conferences. She has twice received the Second Runner-up position of the Group Emotion Recognition Sub-challenge sponsored by ICMI, ACM. Her main research interests include machine learning, computer vision, and affective computing.}
\vspace{4\baselineskip}

\par\noindent
\parbox[t]{\linewidth}{
\noindent\parpic{\includegraphics[height=1.2in,width=1in,clip,keepaspectratio]{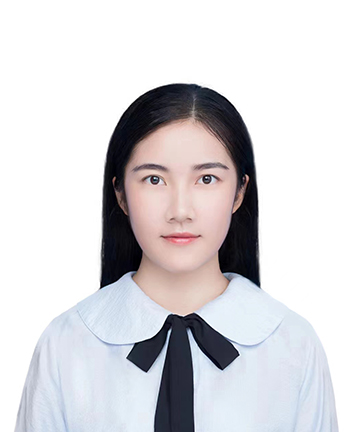}}
\noindent {\bf Yaqi Zhou}\
 is currently a gradutate student at the School of Information and Communication Engineering, Communication University of China. Her research interests include multimodal learning, deep learning and emotion analysis. }
\vspace{4\baselineskip}

\par\noindent
\parbox[t]{\linewidth}{
\noindent\parpic{\includegraphics[height=1.2in,width=1in,clip,keepaspectratio]{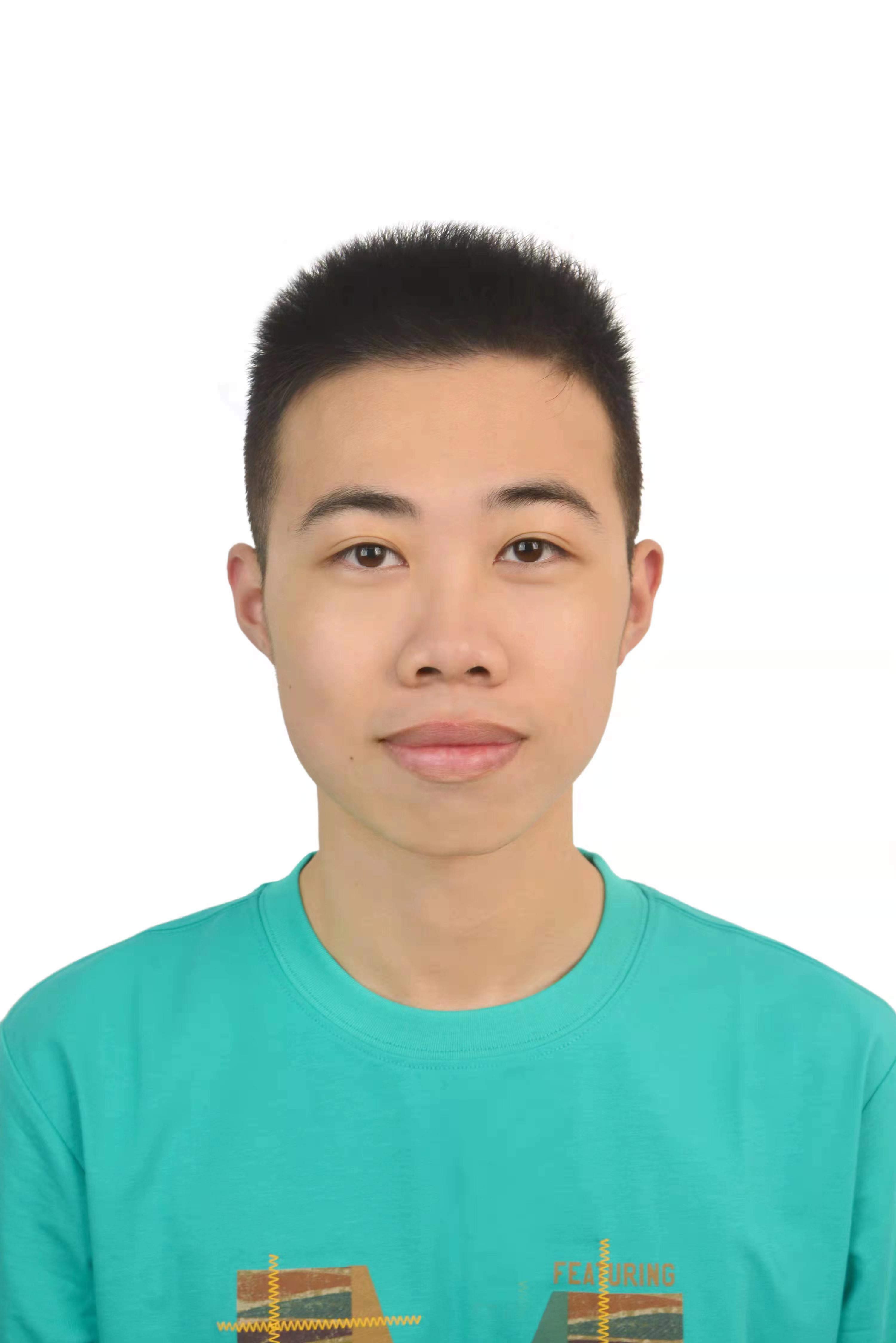}}
\noindent {\bf Longhui Xiao}\
is currently an undergraduate student with interests with the 
School of Data Science and Intelligent 
Media, Communication University of 
China. His research interests include computational sociology, emotion analysis, and disaster management. 
}
\vspace{4\baselineskip}

\par\noindent
\parbox[t]{\linewidth}{
\noindent\parpic{\includegraphics[height=1.2in,width=1in,clip,keepaspectratio]{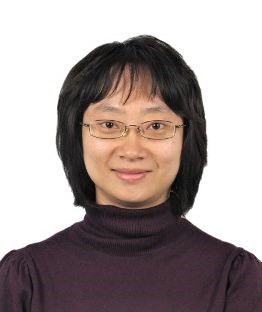}}
\noindent {\bf Yuan zhang}\
received the B.Sc. and 
M.Sc. degrees in electronic 
engineering from Communication 
University of China (CUC), Beijing, 
in 1995 and in 1998, respectively, 
and the Ph.D. degree in computer 
software and theory from the 
Graduate School of the Chinese 
Academy of Sciences, in 2007. She 
was a Visiting Scholar with the Department of Electrical 
and Computer Engineering, University of California, San 
Diego, USA, from September 2010 to August 2011. She is 
currently a Full Professor, the Vice Director of State Key 
Laboratory of Media Convergence and Communication, Communication University of China, Beijing.}
\vspace{4\baselineskip}

\end{document}